\documentclass{PoS}

\usepackage{epsfig}

\def\beq{\begin{equation}}
\def\eeq{\end{equation}}
\def\beqa{\begin{eqnarray}}
\def\eeqa{\end{eqnarray}}

\title{NNLL resummation for $W$-boson production at large $p_T$}

\ShortTitle{NNLL resummation for $W$-boson production at large $p_T$}

\author{\speaker{Nikolaos Kidonakis}%
         \thanks{This material is based upon work supported by the National Science Foundation under Grant No. PHY 1212472.}\\
        Kennesaw State University, USA\\
        E-mail: \email{nkidonak@kennesaw.edu}}

\author{Richard J. Gonsalves\\
        University at Buffalo, The State University of New York, USA\\
        E-mail: \email{phygons@buffalo.edu}}

\abstract{We present new results for $W$-boson production at large transverse momentum at the LHC and the Tevatron. The contribution of soft-gluon corrections is derived from NNLL resummation and added to the exact NLO result.
Numerical results and their uncertainties for the approximate NNLO $W$-boson transverse momentum distributions are derived and compared to recent data from the LHC.}

\FullConference{36th International Conference on High Energy Physics\\
                 July 4-11, 2012\\
                 Melbourne, Australia}

\begin{document}

\section{Introduction}

$W$ hadroproduction is useful in testing the Standard Model and in
estimates of backgrounds to Higgs production and new physics.
The transverse momentum, $p_T$, distribution of the $W$ boson falls
rapidly as $p_T$ increases.
Its behavior at high (and also low) $p_T$ is sensitive to assumptions that
go into perturbative QCD predictions, and to parton distribution functions,
and it provides important input to precision measurements of the $W$ mass.

The partonic channels for this process at LO are
$q(p_a) + g(p_b) \longrightarrow W(Q) + q(p_c)$
and
$q(p_a) + {\bar q}(p_b) \longrightarrow W(Q) + g(p_c)$.
At NLO and NNLO the gluon initiated channels dominate the
cross section at multi-TeV energies.
We define the kinematical variables $s=(p_a+p_b)^2$, $t=(p_a-Q)^2$,
$u=(p_b-Q)^2$, and $s_4=s+t+u-Q^2$. At partonic threshold for a given $p_T$,  
$s_4 \rightarrow 0$.

In this contribution we present both NLO and approximate NNLO results
for the $p_T$ distribution of the $W$-boson at large $p_T$. The NLO calculation
is complete while at NNLO we include soft-gluon corrections that are dominant
near partonic threshold.

\section{NLO $p_T$ distribution of the $W$-boson at the LHC and the Tevatron}

The NLO cross section can be written as
\beqa
E_Q\,\frac{d\hat{\sigma}_{f_af_b{\rightarrow}W(Q)+X}}{d^3Q}&=&
\delta(s_4) \, \alpha_s(\mu_R^2) \, \left[A(s,t,u) \right.
\nonumber \\ && \left.
{}+\alpha_s(\mu_R^2) \, 
B(s,t,u,\mu_R)\right] + \alpha_s^2(\mu_R^2) \, C(s,t,u,s_4,\mu_F)
\eeqa
where the coefficient functions $A$, $B$, and $C$ depend on the parton flavors,
$\mu_F$ is the factorization scale, and $\mu_R$ is the renormalization scale.
For numerical results we set $\mu \equiv \mu_F=\mu_R$.

The coefficient $A(s,t,u)$ arises from the LO processes,
$B(s,t,u,\mu_R)$ is the sum of virtual corrections and of singular terms
${\sim}\delta(s_4)$ in the real radiative corrections, and
$C(s,t,u,s_4,\mu_F)$ is from real emission processes away from $s_4=0$.
The complete NLO analytical results were derived over twenty years ago \cite{AR,gpw}.

\begin{figure}
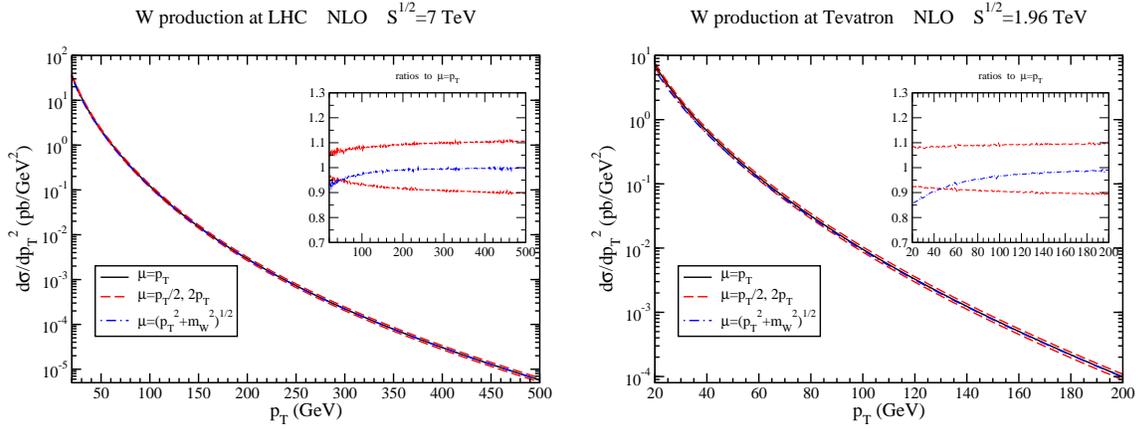

\includegraphics[width=.48\textwidth]{W7lhcNLOmuplot.eps}
\hspace{3mm}
\includegraphics[width=.48\textwidth]{WtevNLOmuplot.eps}
\caption{$W$-boson NLO $p_T$ distributions with various choices of
scale at the LHC at 7 TeV energy (left) and at the Tevatron (right).}
\label{NLO}
\end{figure}

In Fig. \ref{NLO} we show NLO results for $d\sigma/dp_T^2$ at the LHC at 7 TeV energy (left) and at the Tevatron at 1.96 TeV energy (right). We use the MSTW 2008 pdf sets \cite{MSTW}. Results are shown for four different choices of scale: the central choice $\mu=p_T$, its variation by a factor of two, $\mu=p_T/2$, $2p_T$, and also $\mu=(p_T^2+m_W^2)^{1/2}$. The inset plots show ratios of the differential distributions with the last three choices to the central result with $\mu=p_T$.

\section{Approximate NNLO $p_T$ distribution of the $W$-boson at the LHC and the Tevatron}

\begin{figure}
\includegraphics[width=.31\textwidth]{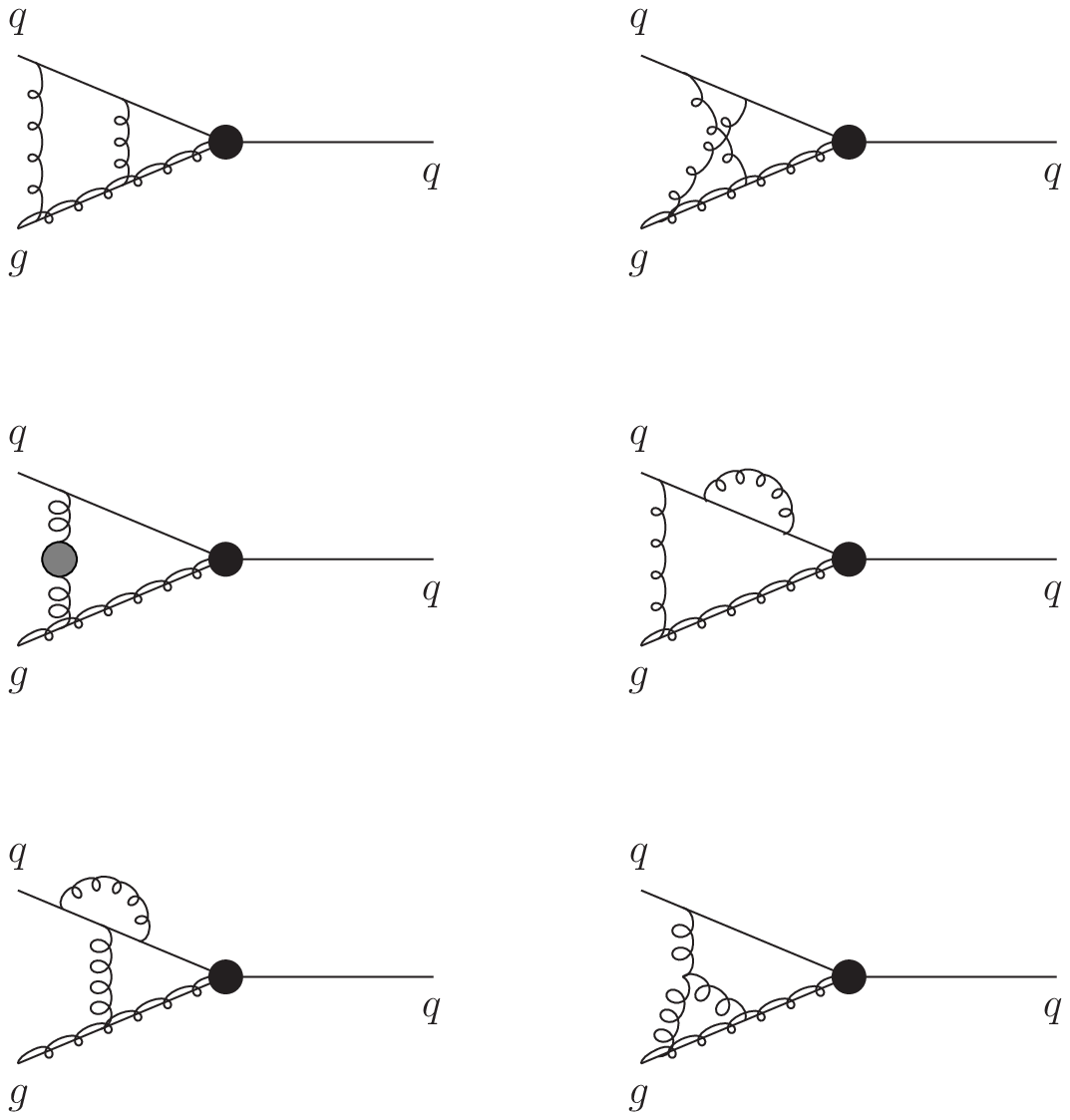}
\hspace{2mm}
\includegraphics[width=.31\textwidth]{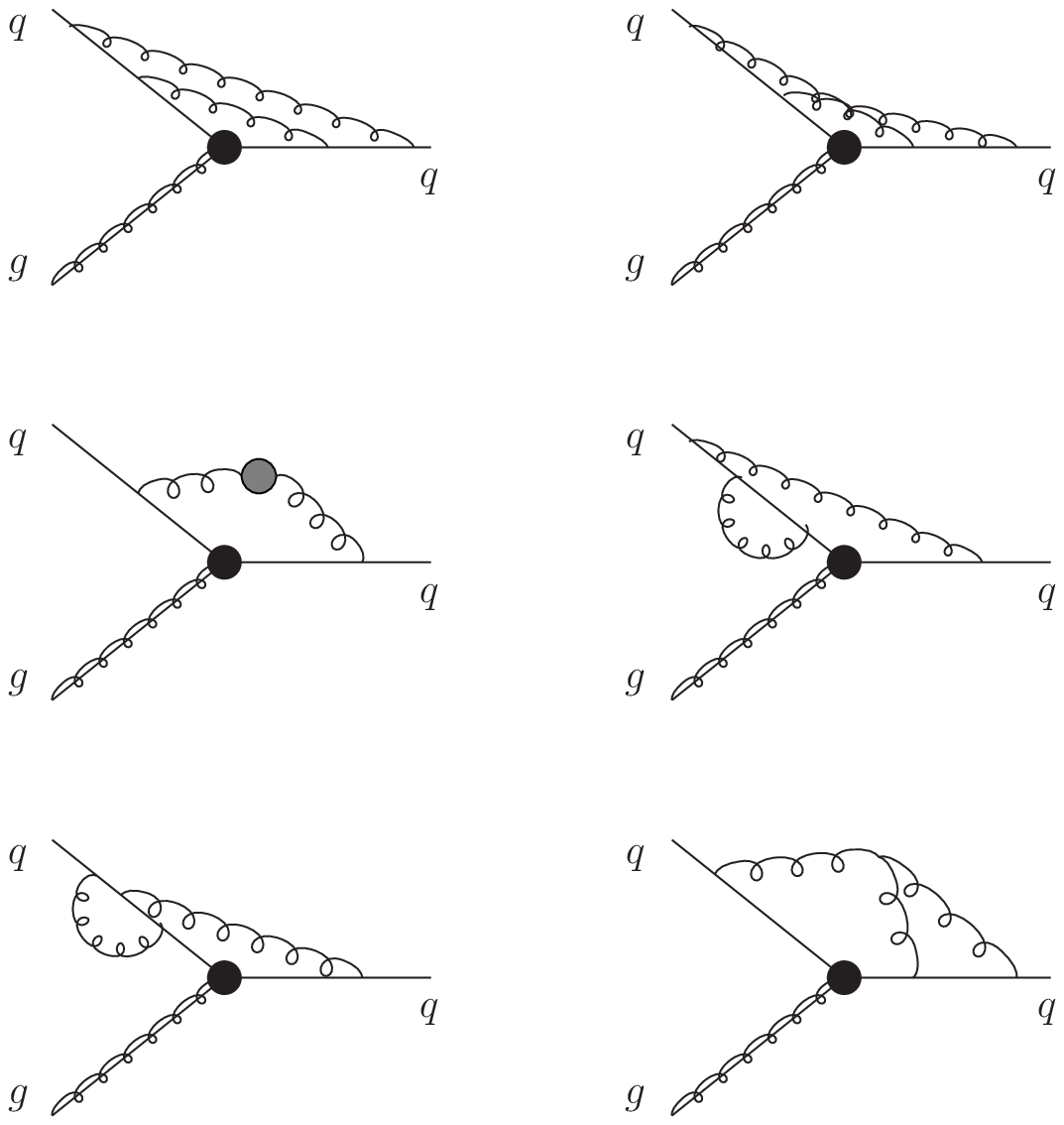}
\hspace{2mm}
\includegraphics[width=.31\textwidth]{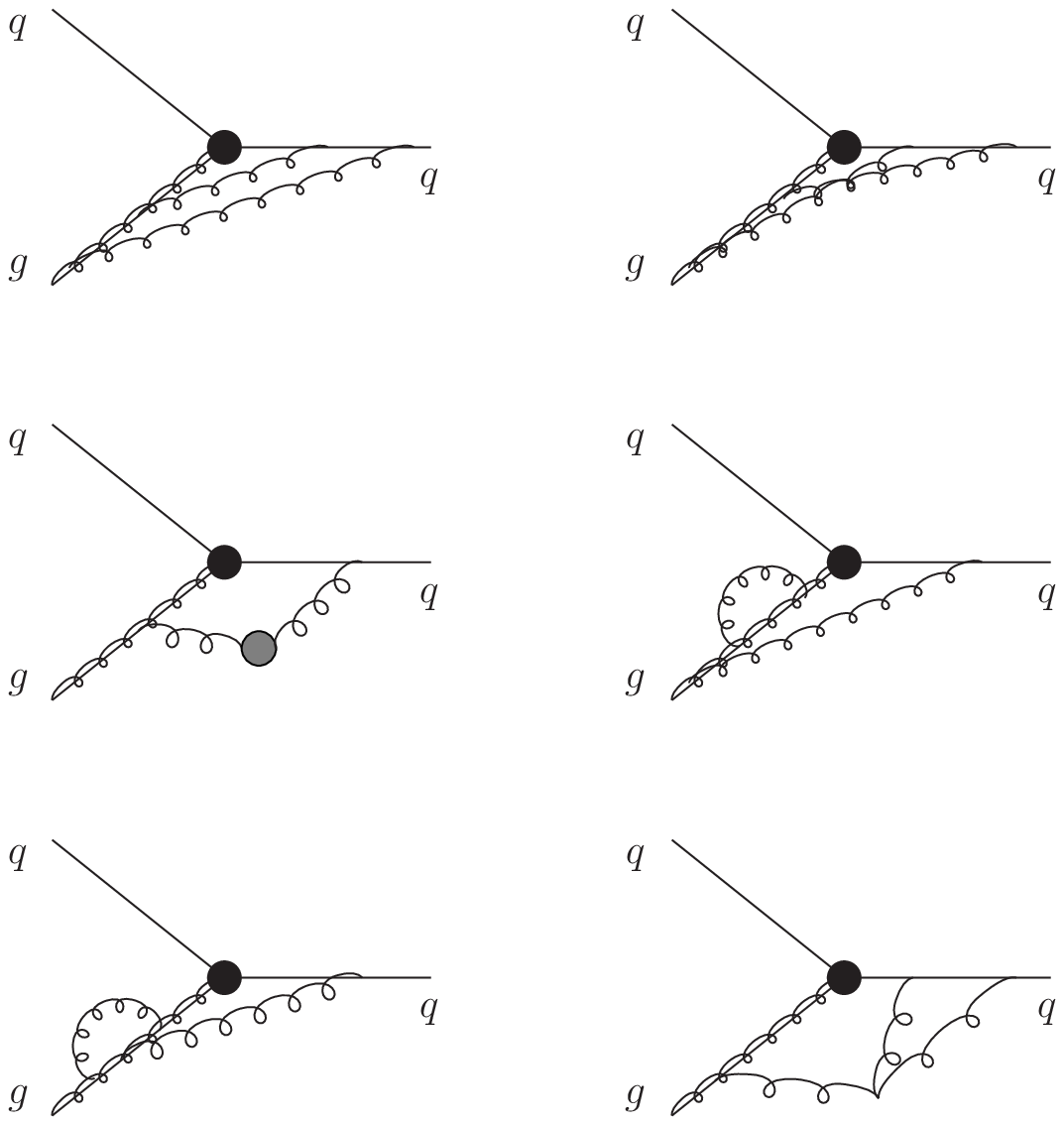}
\caption{Two-loop eikonal diagrams for $qg \rightarrow Wq$.}
\label{Wqg2loop}
\end{figure}

\begin{figure}
\includegraphics[width=.31\textwidth]{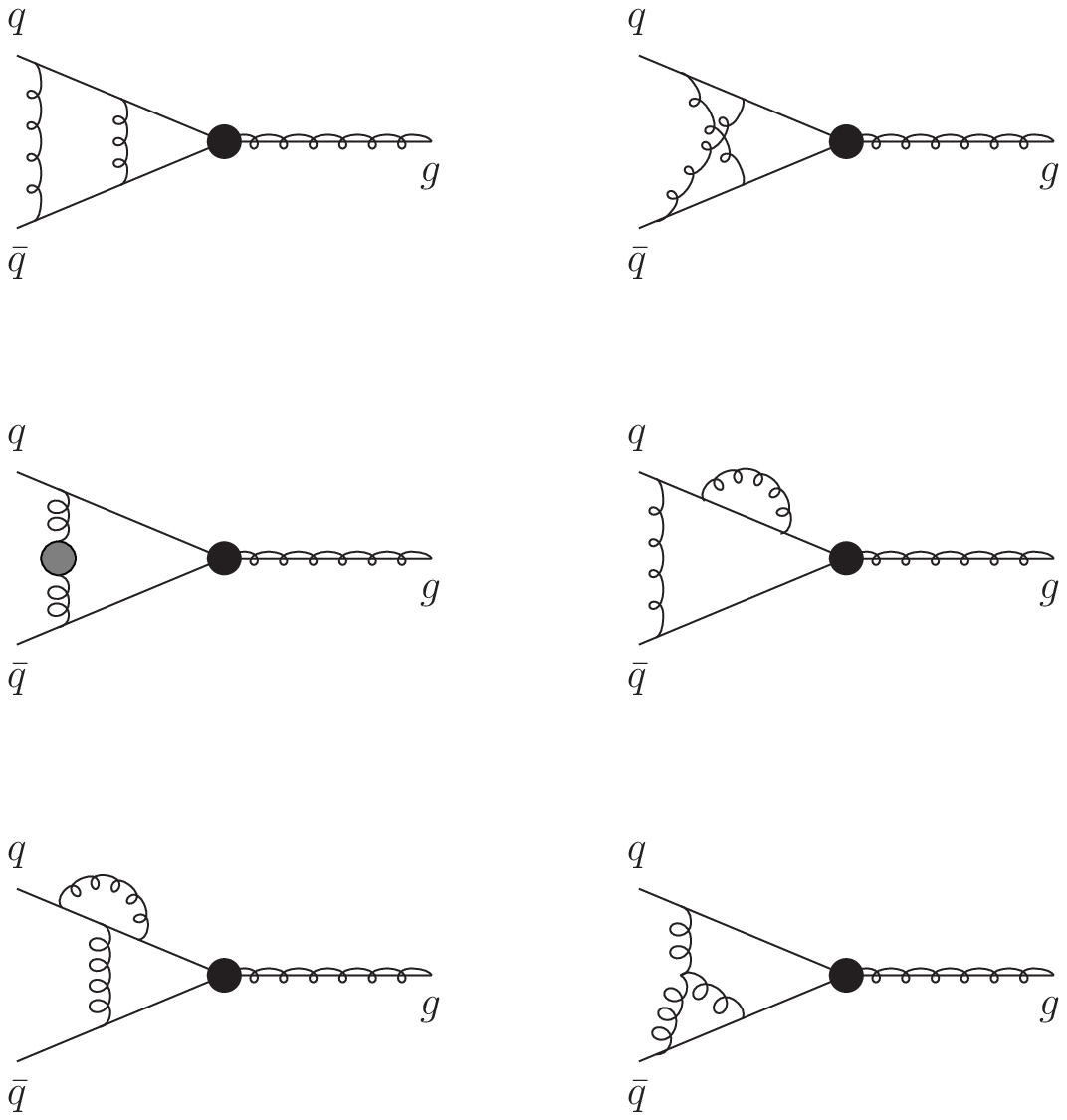}
\hspace{2mm}
\includegraphics[width=.31\textwidth]{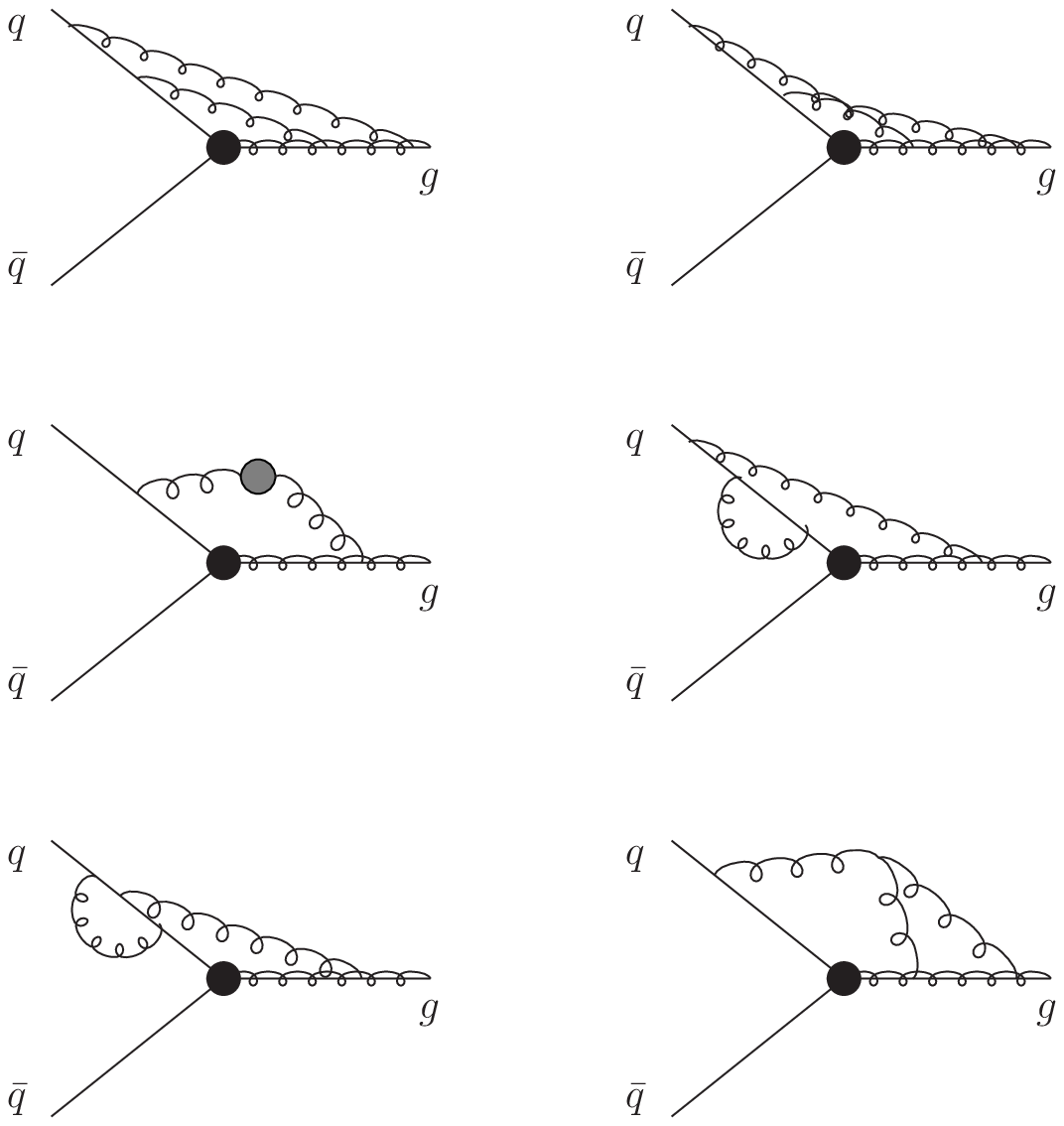}
\hspace{2mm}
\includegraphics[width=.31\textwidth]{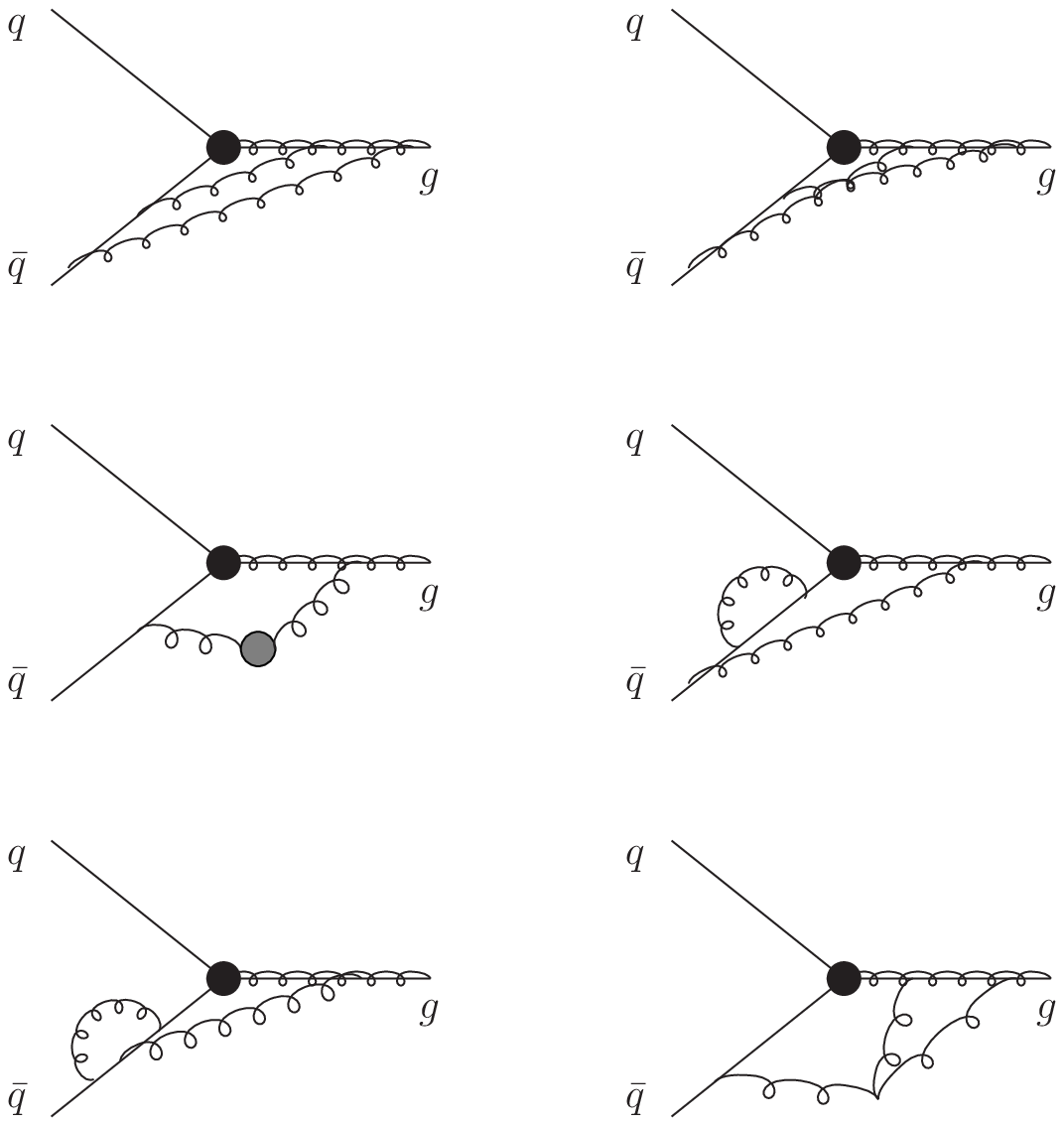}
\caption{Two-loop eikonal diagrams for $q{\bar q} \rightarrow Wg$.}
\label{Wqq2loop}
\end{figure}

The soft-gluon corrections, which appear at each order of perturbation theory starting at NLO, are of the form
$$
{\cal D}_l(s_4)\equiv\left[\frac{\ln^l(s_4/p_T^2)}{s_4}\right]_+ \, .
$$

For the order $\alpha_s^n$ corrections $l\le 2n-1$.
We can formally resum these logarithms for $W$ production at
large $p_T$ to all orders in $\alpha_s$ \cite{NKVD}, and from the expansion
of the resummed cross section derive approximate NNLO results.
Thus, at NLO, the corrections include ${\cal D}_1(s_4)$ and ${\cal D}_0(s_4)$ terms. At NNLO, the corrections include ${\cal D}_3(s_4)$, ${\cal D}_2(s_4)$, ${\cal D}_1(s_4)$, and ${\cal D}_0(s_4)$ terms.

The formalism was previously applied to $W$ production at the Tevatron in
Ref. \cite{NKASV} and at the LHC at 14 TeV energy in Ref. \cite{GKS}.
Those calculations were based on NLL resummation \cite{NKVD}
with additional subleading
terms. Thus at NNLO all ${\cal D}_3(s_4)$, ${\cal D}_2(s_4)$, and ${\cal D}_1(s_4)$ terms were fully determined in \cite{NKASV} but only partial ${\cal D}_0(s_4)$ terms could be calculated.

Recent two-loop results for the soft anomalous dimensions \cite{NKDISDPF} have made possible NNLL resummation and thus the determination of all NNLO ${\cal D}_0(s_4)$ terms \cite{NKRJG}.
These new approximate NNLO analytical results from the expansion of the NNLL resummed differential cross section are then used here to provide numerical predictions \cite{NKRJG, NKRJGDPF} for the $W$-boson $p_T$ distribution at the LHC and the Tevatron.
We note that a different, SCET-based, approach to resummation for the $W$ boson has also appeared in \cite{BLS}.

Soft-gluon resummation follows from the factorization properties of the
cross section and is  performed in moment space. The resummed cross section is
\beqa
{\hat{\sigma}}^{res}(N) &=&
\exp\left[ \sum_i E_i(N_i)\right] \, \exp\left[ E'_j(N')\right]\;
\exp \left[\sum_{i=1,2} 2 \int_{\mu_F}^{\sqrt{s}} \frac{d\mu}{\mu}\;
\gamma_{i/i}\left({\tilde N}_i, \alpha_s(\mu)\right)\right] \;
\nonumber\\ && \times \,
H\left(\alpha_s\right) \;
S \left(\alpha_s\left(\frac{\sqrt{s}}{\tilde N'}\right)\right) \;
\exp \left[\int_{\sqrt{s}}^{{\sqrt{s}}/{\tilde N'}}
\frac{d\mu}{\mu}\; 2\, {\rm Re} \Gamma_S\left(\alpha_s(\mu)\right)\right]
\nonumber
\eeqa
where the first two exponents describe collinear and soft emission from incoming and outgoing partons,  
$H$ is the hard function, $S$ is the soft-gluon function, and 
$\Gamma_S$ is the soft anomalous dimension with perturbative expansion
$$
\Gamma_S=\frac{\alpha_s}{\pi}\Gamma_S^{(1)}
+\frac{\alpha_s^2}{\pi^2}\Gamma_S^{(2)}
+\cdots
$$

\begin{figure}
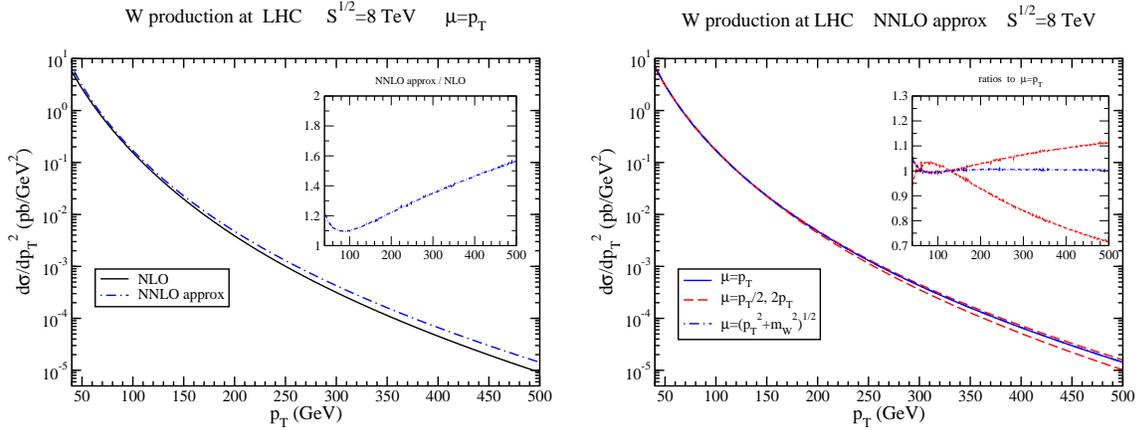

\includegraphics[width=.48\textwidth]{W8lhcnnloratioplot.eps}
\hspace{3mm}
\includegraphics[width=.48\textwidth]{W8lhcptnnlomuplot.eps}
\caption{$W$-boson approximate NNLO $p_T$ distributions with various choices of
scale at the LHC at 8 TeV.}
\label{NNLO8}
\end{figure}

The two-loop soft anomalous dimension, $\Gamma_S^{(2)}$, is determined
from the UV poles of dimensionally-regularized integrals for the two-loop
eikonal diagrams for $qg \rightarrow Wq$, shown in Fig. \ref{Wqg2loop},
and $q{\bar q} \rightarrow Wg$, shown in Fig. \ref{Wqq2loop}.

The analytical expressions for the soft anomalous dimensions
at one and two loops are given below.
For $qg\rightarrow Wq$
$$
\Gamma_{S,\, qg\rightarrow Wq}^{(1)}=C_F \ln\left(\frac{-u}{s}\right)
+\frac{C_A}{2} \ln\left(\frac{t}{u}\right)
$$
$$
\Gamma_{S,\, qg \rightarrow Wq}^{(2)}=\frac{K}{2} \Gamma_{S,\, qg \rightarrow Wq}^{(1)} \, .
$$
For $q {\bar q}\rightarrow Wg$
$$
\Gamma_{S,\, q{\bar q}\rightarrow Wg}^{(1)}=\frac{C_A}{2} \ln\left(\frac{tu}{s^2}\right)
$$
$$
\Gamma_{S,\, q{\bar q} \rightarrow Wg}^{(2)}=\frac{K}{2} \Gamma_{S,\, q{\bar q} \rightarrow Wg}^{(1)} \, .
$$

Our numerical results for the NNLO approximate $p_T$ distribution of the
$W$ boson at the LHC at 8 TeV energy is shown in Fig. \ref{NNLO8}.
Again, we use the MSTW 2008 pdf sets \cite{MSTW}.
The left plot contrasts the NLO and approximate NNLO results and the inset
plot shows the ratio NNLO approximate / NLO. The NNLO soft-gluon
contributions increasingly enhance the NLO result with increasing $p_T$.
The right plot displays the approximate NNLO $d\sigma/dp_T^2$  for four
different choices of scale: the central choice $\mu=p_T$, its variation by
a factor of two, $\mu=p_T/2$, $2p_T$, and also $\mu=(p_T^2+m_W^2)^{1/2}$.
The inset plot shows ratios of the $p_T$ distributions with the last three
scale choices to the central result with $\mu=p_T$.

\begin{figure}
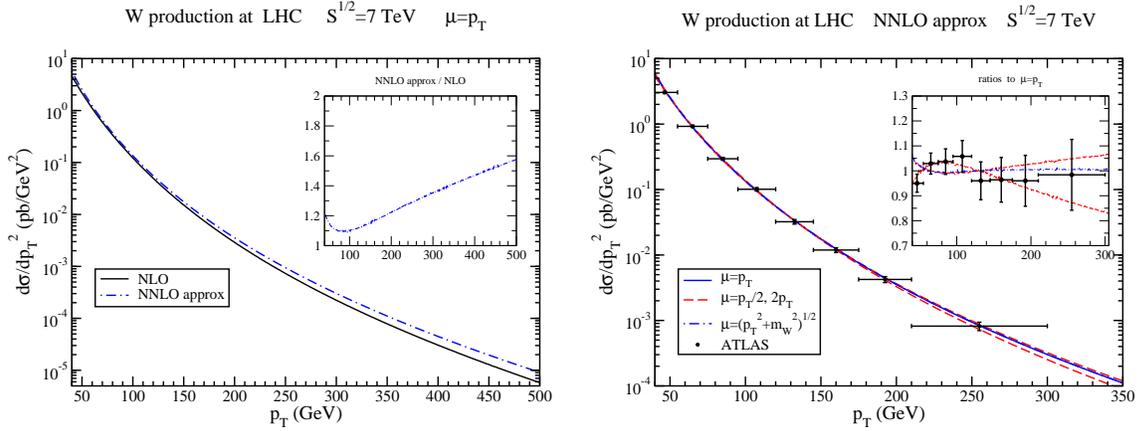

\includegraphics[width=.48\textwidth]{W7lhcnnloratioplot.eps}
\hspace{3mm}
\includegraphics[width=.48\textwidth]{W7lhcptnnlomuATLASplot.eps}
\caption{$W$-boson approximate NNLO $p_T$ distributions with various choices of
scale at the LHC at 7 TeV.}
\label{NNLO7}
\end{figure}

Figure \ref{NNLO7} shows results for 7 TeV LHC energy.
The left plot again contrasts NLO and approximate NNLO results for
$d\sigma/dp_T^2$. Their ratio
is very similar to that at 8 TeV shown previously. The right plot shows
$d\sigma/dp_T^2$ for various scale choices and compared with recent ATLAS
results \cite{ATLAS}. The agreement between our theoretical predictions
and the ATLAS data is very good.

\begin{figure}
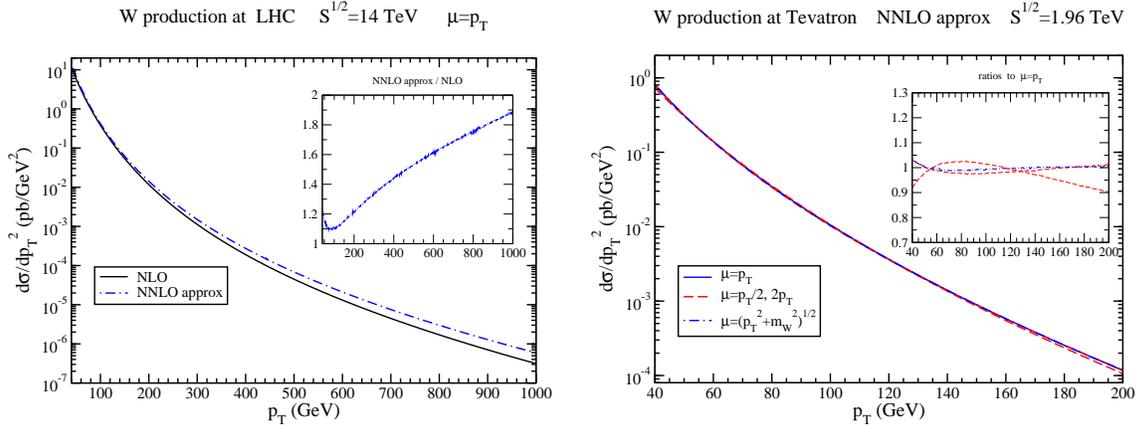

\includegraphics[width=.48\textwidth]{W14lhcnnloratioplot.eps}
\hspace{3mm}
\includegraphics[width=.48\textwidth]{Wtevptnnlomuplot.eps}
\caption{$W$-boson approximate NNLO $p_T$ distributions at 14 TeV LHC (left)
and at the Tevatron (right).}
\label{NNLOlhctev}
\end{figure}

Finally, Fig. \ref{NNLOlhctev} displays related results for the $W$-boson
$p_T$ distribution at the LHC at 14 TeV energy(left plot) and at the Tevatron
(right plot).

More analytical and numerical results can be found in Ref. \cite{NKRJG}
(see also \cite{NKDISDPF,NKRJGDPF}).

\section{Conclusions}

We have calculated the $W$-boson transverse momentum distribution for
$W$ production at large $p_T$. We presented NLO results as well as
approximate NNLO results from NNLL resummation using two-loop soft-gluon
threshold corrections. The higher-order corrections are important for
greater theoretical accuracy.

The approximate NNLO $p_T$ distributions of the $W$ have been calculated
for both LHC and Tevatron energies and are in good agreement with recent
ATLAS data from the LHC at 7 TeV.
It will be instructive to compare our results with 8 TeV data from
ATLAS and CMS.

Our results are also relevant at Tevatron energies.  We note that while
the D0 and CDF Collaborations have published their measurements of the
$W$ boson $p_T$ distributions based on Run I data, corresponding results
for Run II data have not been published to date.
Event rates for this process are expected to increase dramatically when
the LHC resumes operation.
Comparison of reliable and accurate perturbative predictions with experimental
data at these widely separated energy regimes around 2, 8, and 14 TeV could
provide important insight into problems involving parton distribution
functions, scale dependence, and resummation of large logarithms.

\end{document}